\documentclass[prd,twocolumn,showpacs,preprintnumbers,superscriptaddress,floatfix,nofootinbib]{revtex4-1}
\usepackage{subfigure}
\usepackage{amsfonts}
\usepackage{graphicx,,booktabs}
\usepackage{amssymb}
\usepackage{amsmath,txfonts,ulem}
\usepackage{graphicx}
\usepackage{dcolumn}
\usepackage{color}
\usepackage{bm}
\usepackage{multirow}
\usepackage{ulem}
\usepackage[colorlinks,
            citecolor=blue,
            anchorcolor=green,
            menucolor=orange,
            linkcolor=red,
            filecolor=red,
            runcolor=pink,
            urlcolor=blue,
            frenchlinks=red]{hyperref}

\allowdisplaybreaks[4]

\begin{document}

\title{Analysis of the interaction between $\phi$ meson and nucleus}

	\author{Xiao-Yun Wang}
	\email{Corresponding author: xywang@lut.edu.cn}
	\affiliation{Department of physics, Lanzhou University of Technology,
		Lanzhou 730050, China}
	\affiliation{Lanzhou Center for Theoretical Physics, Key Laboratory of Theoretical Physics of Gansu Province, Lanzhou University, Lanzhou, Gansu 730000, China}
	
	\author{Chen Dong}
	\email{dongphysics@yeah.net}
	\affiliation{Department of physics, Lanzhou University of Technology,
		Lanzhou 730050, China}
	
	\author{Quanjin Wang}
	\affiliation{Department of physics, Lanzhou University of Technology,
		Lanzhou 730050, China}

\begin{abstract}
In this work, we systematically study the $\phi$ meson and nucleus interaction by analyzing and fitting the cross sections of $\gamma N$$\rightarrow \phi$$N$ ($N$ represent the nucleus) reactions near the threshold. With the help of vector meson dominant model, the distribution of $\phi$-$N$ scattering length as a function of energy is presented, and the results show that there is a slight increase in scattering length with increasing energy.
Based on this, the average scattering length of $\phi$-proton is obtained as $0.10\pm0.01$ fm by combining experimental data and theoretical models. Moreover, the average scattering length of $\phi$-deuteron interaction is derived to be $0.014\pm0.002$ fm for the first time.
Further, the effect of the momentum transfer $|t_{min}|$ on the $\phi$-$N$ scattering length at the threshold is discussed. The obtained results not only provide important theoretical information for a more comprehensive and accurate study of the $\phi$-$N$ scattering length, but also provide a basis for future experimental measurements of $\phi$ meson production.

\end{abstract}


\maketitle

\section{Introduction}
Vector mesons in the low energy photoproduction play an essential role in understanding vector mesons-proton ($V$-$p$) interactions \cite{Titov:2007xb}. The most apparent manifestation is the scattering length of $V$-$p$ interaction. Researchers now calculate the scattering length of $V$-$p$ interaction by establishing theoretical models or analyzing experimental data, such as $\rho$-$p$, $\omega$-$p$, $\phi$-$p$, $J/\psi$-$p$, $\psi(2S)$-$p$, $\Upsilon$-$p$ \cite{Koike:1996ga,Chang:2007fc,Titov:2007xb,Strakovsky:2014wja,Strakovsky:2019bev,Strakovsky:2020uqs,Pentchev:2020kao,Wang:2022xpw,Wang:2022zwz,Strakovsky:2021vyk}. As a conventional $s\bar{s}$ state, $\phi$ has been extensively concerned. In order to study the internal structure of the proton, some researchers extract the mass radius of the proton from the cross section of $\phi$ photoproduction at the near-threshold \cite{Wang:2021ujy,Wang:2022uch}. Others have moved on to the scattering length $\phi$-$p$ interaction. The earliest can be traced back to 1997, Yuji Koike et al. \cite{Koike:1996ga} used the QCD sum rule to analyze the scattering length of spin-isospin average $\rho$-$p$, $\omega$-$p$ and $\phi$-$p$. By the low energy limit constraint of the forward scattering amplitude of the vector-current nucleon, $\alpha_{\rho p}=-0.47\pm0.05$ fm, $\alpha_{\omega p}=-0.41\pm0.05$ fm and $\alpha_{\phi p}=-0.15\pm0.02$ fm were obtained.  In 2000, H. Gao et al. \cite{Gao:2000az} analyzed the $\phi$-$N$ bound state through QCD van der Waals attractive potential and believed that the $\phi$-$N$ bound state can help detect the strangeness content of nucleons. In 2007, the LEPS Collaboration \cite{Chang:2007fc} obtained $\alpha_{\phi p}=-0.15$ fm from the differential experimental cross sections of $\gamma p$$\rightarrow$$\phi p$ at the near-threshold for the first time. In the same period, A. I. Titov et al. \cite{Titov:2007xb} related the differential cross section of $\phi$ photoproduction to scattering length by establishing vector meson dominant (VMD) model. It indicates that the differential cross section of $\phi$ at the threshold is finite, and its behaviour is crucial for the QCD-inspired $\phi$-$p$ interaction model. Since then, the study of the scattering length of $\phi$-$p$ interaction has fallen silent.

In recent years, the investigation of the scattering length of $V$-$p$ interaction has been revived due to the accumulation of the experimental data of vector mesons photoproduction. In 2014, Igor I. Strakovsky et al. \cite{Strakovsky:2014wja} used odd power polynomials to fit the total experimental cross section of $\omega$ photoproduction and obtained $|\alpha_{\omega p}|=0.81\pm0.41$ fm. In 2020, they \cite{Strakovsky:2019bev,Strakovsky:2020uqs} combined the total experiment cross section of the vector meson photoproduction with the VMD model. $|\alpha_{J/\psi p}| =3.08 \pm 0.55$ am and $|\alpha_{\phi p}|=0.063\pm0.01$ fm were obtained by fitting the odd power polynomial with the latest photoproduction data. Later, the same researchers \cite{Pentchev:2020kao} combined the differential cross section of $J/\psi$ photoproduction with the scattering length based on the VMD model. $|\alpha_{J/\psi p}|=3.83\pm 0.98$ am was obtained utilising  GlueX \cite{GlueX:2019mkq} and SLAC \cite{Gryniuk:2016mpk} data. In our previous work \cite{Wang:2022xpw,Wang:2022zwz}, the same method was used to calculate the average scattering lengths $|\alpha_{J/\psi p}|=3.85\pm0.96$ am, $|\alpha_{\psi(2s) p}|=1.31\pm0.92$ am and $|\alpha_{\rho p}|=0.29\pm0.07$ fm at the near-threshold. In summary of the above work, we find that the mass of vector meson is inversely proportional to the scattering length of $V$-$p$ interaction, $|\alpha_{\omega p}|>|\alpha_{\phi p}|>|\alpha_{J/\psi p}|>|\alpha_{\psi(2S) p}|>|\alpha_{\Upsilon p}|$ \cite{Wang:2022zwz}. Here, $\rho$ is temporarily excluded due to the particular situation. 

For the scattering length of $\phi$-$p$ interaction, many groups have also performed calculations and measurements. The scattering length $|\alpha_{\phi p} |\simeq2.37$ was obtained by analyzing the QCD van der Waals potential \cite{Gao:2000az}. And in 2021, the real part of the scattering length of $\phi$-$p$ was found to be $0.85\pm0.34$ fm by $pp$ collision in ALICE Collaboration \cite{ALICE:2021cpv}. Note that these results go beyond the scattering length of the $\phi$-$p$ extracted from the vector meson photoproduction data. Moreover, the scattering length of $\phi$-$p$ obtained by the VMD model is basically extracted from the single data of vector meson photoproduction \cite{Koike:1996ga,Gao:2000az,Chang:2007fc,Titov:2007xb}. Considering that the extraction of the $\phi$-$p$ scattering length from a single experimental data point may have some uncertainty, it is necessary to give the distribution of the $\phi$-$p$ scattering length with energy at the threshold.
Therefore, in this work, the two gluon exchange model \cite{Wang:2022uch} and the pomeron model \cite{Laget:1994ba,Sibirtsev:2004ca} are established to predict the cross section of $\phi$ photoproduction. With the VMD model, the scattering length $|\alpha_{\phi p}|$ is connected to the photoproduction cross section and expressed as a function of $R$. Combining with the $|\alpha_{\phi p}|$ extracted directly from the experimental data \cite{Strakovsky:2020uqs,Dey:2014tfa,LEPS:2005hax}, the value of $|\alpha_{\phi p}|$ at the near threshold $R$ can be obtained. Here, $R$ is the ratio of the final momentum $|{\bf p}_3|$ to the initial momentum $|{\bf p}_1|$, which is directly proportional to the center of mass energy $W$.

At present, the research on the scattering length of vector mesons interacting with deuteron or helium nuclei composed of multiple nucleons is still very limited. Considering the validity of VMD model in describing the photoproduction process of vector meson and nucleus coherence \cite{VMD,Gell-Mann:1961jim,Harari:1966ms,Stodolsky:1966am,Dar:1968sla}, this paper will analyze and calculate the scattering length of $\phi$-deuteron ($d$). Fortunately, in 2007, the SLEP Collaboration \cite{Chang:2007fc} newly measured the coherent $\phi$ photoproduction from the deuterons at $E$=$1.5$-$2.4$ GeV with a forward angle and linearly polarized beam, which provided important experimental data for our study of the scattering length of $\phi$-$d$.
The findings of this paper not only contribute to our in-depth understanding of the $\phi$-$N$ interaction, but also provide a theoretical basis for future experimental measurements at JLab or EIC facilities \cite{Accardi:2012qut,Anderle:2021wcy}.

The outline of this work is organized as follows. The expression of scattering length correlation, the two gluon exchange and the pomeron models are described in Sec. \ref{sec2}. The results of scattering length of $\phi$-$N$ coherence are presented in Sec. \ref{sec3}. A simple summary in Sec. \ref{sec4}.
\section{Formalism} \label{sec2}
\subsection{Scattering Length}
The  total cross section of $\phi$ photoproduction at the near-threshold is related to the scattering length of $\phi$-N interaction ($|\alpha_{\phi N}|$) with the Vector Meson Dominance (VMD) model \cite{Pentchev:2020kao},
\begin{equation}
\begin{aligned}
\left.\sigma ^{\gamma N \rightarrow \phi N}\right|_{t h r}(R) &=\left.\frac{\left|\mathbf{\bf p}_3\right|}{\left|\mathbf{\bf p}_1\right|} \cdot \frac{4 \alpha_{em} \pi^{2}}{g_{\phi}^{2}} \cdot \frac{d \sigma^{\phi N \rightarrow \phi N}}{d \Omega}\right|_{\text {thr}} \\
&=R \cdot \frac{4 \alpha_{em} \pi^{2}}{g_{\phi}^{2}} \cdot\left|\alpha_{\phi N}\right|^{2},
\end{aligned} \label{eq:1}
\end{equation}
with the VMD coupling constant $g_{\phi}$,
\begin{equation}
    g_{\phi}=\sqrt{\frac{\pi \alpha^{2}_{em}m_{\phi}}{3\Gamma_{\phi \rightarrow e^{+}e^{-}}}}, \label{eq:2}
\end{equation}
where $N$ represent the nucleus, $\alpha_{em}$ is the fine coupling constant, $R$ is the ratio of 
the final momentum $|{\bf p}_3|$ to the initial momentum $|{\bf p}_1|$  and the $\Gamma_{e^{+}e^{-}}=1.27$ keV is the lepton decay width from Ref. \cite{Strakovsky:2020uqs}. With Eq. (\ref{eq:1}) and (\ref{eq:2}), the scattering length $|\alpha_{\phi N}|$ is
\begin{equation}
    |\alpha_{\phi N}|=\frac{g_{\phi}}{2 \pi}\sqrt{\frac{\sigma^{\gamma N\rightarrow \phi N}}{\alpha_{em} R}}. \label{eq:3}
\end{equation}

In the center of mass frame, the momentum of initial and final  are
\begin{equation}
    \left|\mathbf{\bf p}_1\right|=\frac{1}{2 W} \sqrt{W^{4}-2\left(m_{1}^{2}+m_{2}^{2}\right) W^{2}+\left(m_{1}^{2}-m_{2}^{2}\right)^{2}}, \label{eq:4}
\end{equation}
\begin{equation}
    \left|\mathbf{\bf p}_3\right|=\frac{1}{2 W} \sqrt{W^{4}-2\left(m_{3}^{2}+m_{4}^{2}\right) W^{2}+\left(m_{3}^{2}-m_{4}^{2}\right)^{2}}, \label{eq:5}
\end{equation}
where the $W$ is the center of mass energy for the $\gamma p$ collision.

The total cross section can be obtained by differential cross section integration from the four momentum $t_{min}(W)$ to $t_{max}(W)$, which can be given as 
\begin{equation}
    \sigma^{\gamma N \rightarrow \phi N}=\int_{t_{min}(W)}^{t_{max}(W)}\frac{d \sigma^{\gamma N \rightarrow \phi N}}{d t} d t,\label{eq:6}
\end{equation}
with
\begin{equation}
t_{\max }\left(t_{\min }\right)=m_{1}^{2}+m_{3}^{2}-2 E_{1} E_{3} \pm 2\left|\mathbf{p}_{1}\right|\left|\mathbf{p}_{3}\right|,
\end{equation}
where $E_{i}=\sqrt{|{\bf p}_i|^{2}+m_{i}^{2}}$ (i=1, 3).
As the center of mass energy $W$ approaches the threshold, $t_{min}$ approaches $t_{max}$. Therefore, rewritten Eq. (\ref{eq:6}) as
\begin{equation}
\left.\sigma\right|_{t h r}=\left.4\left|\mathbf{p}_{1}\right| \cdot\left|\mathbf{p}_{3}\right| \frac{d \sigma}{d t}\right|_{t h r}\label{eq:7},
\end{equation}
where $|t_{max} - t_{min}|= \Delta t=4 |{\bf p}_3||{\bf p}_1|$ \cite{Pentchev:2020kao}. According to Eq. (\ref{eq:1}), one can express the relation between the scattering length $|\alpha_{\phi N}|$ and the differential cross section of $\phi$ photoproduction at the near-threshold
\begin{equation}
    |\alpha_{\phi N}|=\frac{|{\bf p}_1|g_{\phi}}{\pi}\sqrt{\frac{1}{\alpha_{em}}\frac{d \sigma^{\gamma N \rightarrow \phi N}}{d t}}, \label{eq:8}
\end{equation}
In keeping with Eq. (\ref{eq:3}), the above formula is expressed as a function of $R$,
\begin{equation}
    |\alpha_{\phi N}|=\frac{|{\bf p}_3|g_{\phi}}{R \pi}\sqrt{\frac{1}{\alpha_{em}}\frac{d \sigma^{\gamma N \rightarrow \phi N}}{d t}}. \label{eq:9}
\end{equation}

\subsection{Two gluon exchange model}
\begin{figure}[htbp]
	\begin{center}
		\includegraphics[scale=0.36]{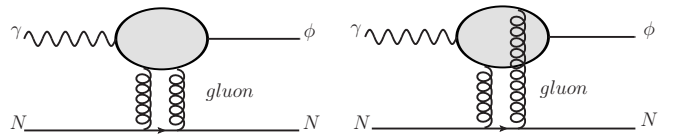}
		\caption{ The Feynman diagrams of the two gluon exchange
model for $\phi$ photoproduction. }  \label{two gluon}
	\end{center}
\end{figure}
The photons split into a pair of dipoles, which exchange two gluons to scatter protons and finally form meson $\phi$ in the two gluon exchange model shown in Fig. \ref{two gluon}. In the lowest order perturbation QCD, the differential cross section obtained by $\phi$ photoproduction amplitude \cite{Sibirtsev:2004ca},
\begin{equation}
\frac{d \sigma}{d t}=\frac{\pi^{3} \Gamma_{e^{+} e^{-}} \alpha_{s}}{6 \alpha m_{s}^{5}}\left[x g\left(x, m_{\phi}^{2}\right)\right]^{2} \exp \left(b_{0} t\right),
\label{eq:10}
\end{equation}
where the QCD coupling constant $\alpha_{s}=0.701$ is from Ref. \cite{Kou:2021bez}, $m_{s}$ is the mass of the strange quark, $m_{\phi}=1.019$ GeV is the mass of meson $\phi$ and $b_{0}$ is the slope. 
$xg(x,m_{\phi}^{2})=A_{0}x^{A_{1}}(1-x)^{A_{2}}$ is the parameterized gluon distribution function, $A_{0}$, $A_{1}$ and $A_{2}$ are free parameters. The total cross section of $\phi$ photoproduction can be obtained by integrating the Eq. (\ref{eq:10}) from $t_{min}(W)\rightarrow t_{max}(W)$
\begin{equation}
    \sigma^{\gamma p \rightarrow \phi p}=\int_{t_{min}(W)}^{t_{max}(W)}\frac{d \sigma^{\gamma p \rightarrow \phi p}}{d t}d t  \label{eq:11}
\end{equation}

\subsection{Pomeron model}
\begin{figure}[htbp]
	\begin{center}
		\includegraphics[scale=0.4]{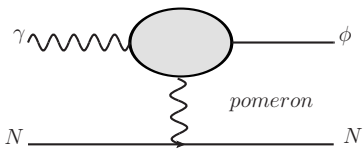}
		\caption{ The Feynman diagram of the pomeron model for $\phi$ photoproduction.  }  \label{pomeron}
	\end{center}
\end{figure}
Fig. \ref{pomeron} represents the $\gamma p$$\rightarrow$$\phi p$ reaction process of the pomeron model \cite{Laget:1994ba,Sibirtsev:2004ca}, the differential cross section of $\phi$ photoproduction is written as,
\begin{equation}
\frac{\mathrm{d} \sigma}{\mathrm{d} t}=\frac{81 m_{\phi}^{3} \beta^{4} \mu_{0}^{4} \Gamma_{\mathrm{e}^{+} \mathrm{e}^{-}}}{\pi \alpha_{em} }(\frac{s}{s_{0}})^{2\alpha(t)-2}F_{1}(t),
\end{equation}
with
\begin{equation}
   F_{1}(t)= \left(\frac{F(t)}{\left(Q^{2}+m_{\phi}^{2}-t\right)\left(Q^{2}+2 \mu_{0}^{2}+m_{\phi}^{2}-t\right)}\right)^{2},
\end{equation}
where $Q^{2}$ is the squared of the virtual photon, $s_0=4$ GeV$^2$, $\alpha_{em}$ is the fine coupling constant and $\mu^{2}=1.1$ GeV$^{2}$. $F(t)$ is the form factor,
\begin{equation}
F(t)=\frac{4 m_{\mathrm{N}}^{2}-2.8 t}{\left(4 m_{\mathrm{N}}^{2}-t\right)(1-t / 0.7)^{2}},
\end{equation}
where $m_{N}$ is the mass of proton or deuteron. And the Regge trajectory $\alpha(t)=1.08+0.25t$ can be obtained from Ref. \cite{Landshoff:1990kj}. 

In our previous work \cite{Wang:2022uch}, $\beta^{2}=4$ GeV$^{-2}$ was corrected by the nucleon and the nucleon scattering amplitude from Ref. \cite{Landshoff:1990kj}. In this work, $\beta$ as free parameter and obtained by fitting the experimental data of $\phi$ photoproduction.

\section{RESULTS}\label{sec3}
\subsection{The scattering length of $\phi$-$p$ interaction}
\begin{figure}[htbp]
	\begin{center}
		\includegraphics[scale=0.4]{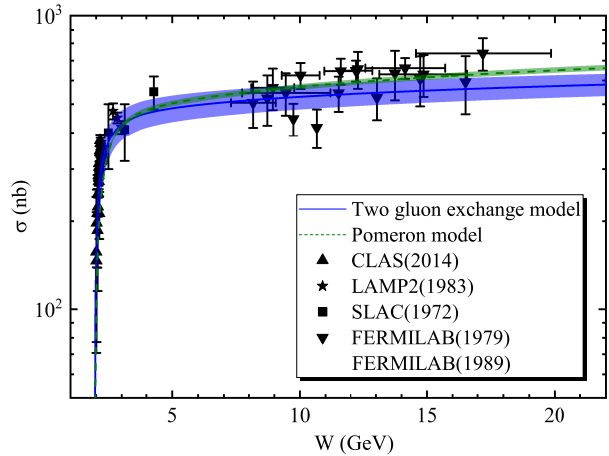}
		\caption{ The total cross section of $\phi$ photoproduction as a function
        of center of mass energy $W$. The solid-line (blue) and dashed line (olive-green)
        are for the two gluon exchange model and pomeron model, respectively. }  \label{fig:tot}
	\end{center}
\end{figure}

In our previous work \cite{Wang:2022uch}, the two gluon exchange model was established to effectively extract the mass radius of the proton, which is extremely proximate to that extracted directly from the CLAS \cite{Dey:2014tfa} and LEPS \cite{LEPS:2005hax} data. The free parameters, $A_0$, $A_1$, $A_2$, and $b_0$, contained in this model were obtained by global fitting of the total \cite{Strakovsky:2020uqs,Barber:1981fj,Ballam:1972eq,Egloff:1979mg,Busenitz:1989gq,Owens:2012bv} and differential \cite{Dey:2014tfa,LEPS:2005hax} cross sections of $\phi$ photoproduction data. In addition, the pomeron model was introduced as an auxiliary model in contrast to the two gluon exchange model. In this work, the two gluon exchange model is still used to calculate the scattering length of $\phi$-$p$ interaction. In addition, $\beta$ in the pomeron model is considered a free parameter obtained by fitting the total \cite{Strakovsky:2020uqs,Barber:1981fj,Ballam:1972eq,Egloff:1979mg,Busenitz:1989gq,Owens:2012bv} and differential \cite{Dey:2014tfa,LEPS:2005hax} experimental cross sections data. The parameters present in the two gluon exchange and pomeron models are shown in Tab. \ref{tab:table3}. 
\begin{figure}[htbp]
	\begin{center}
		\includegraphics[scale=0.43]{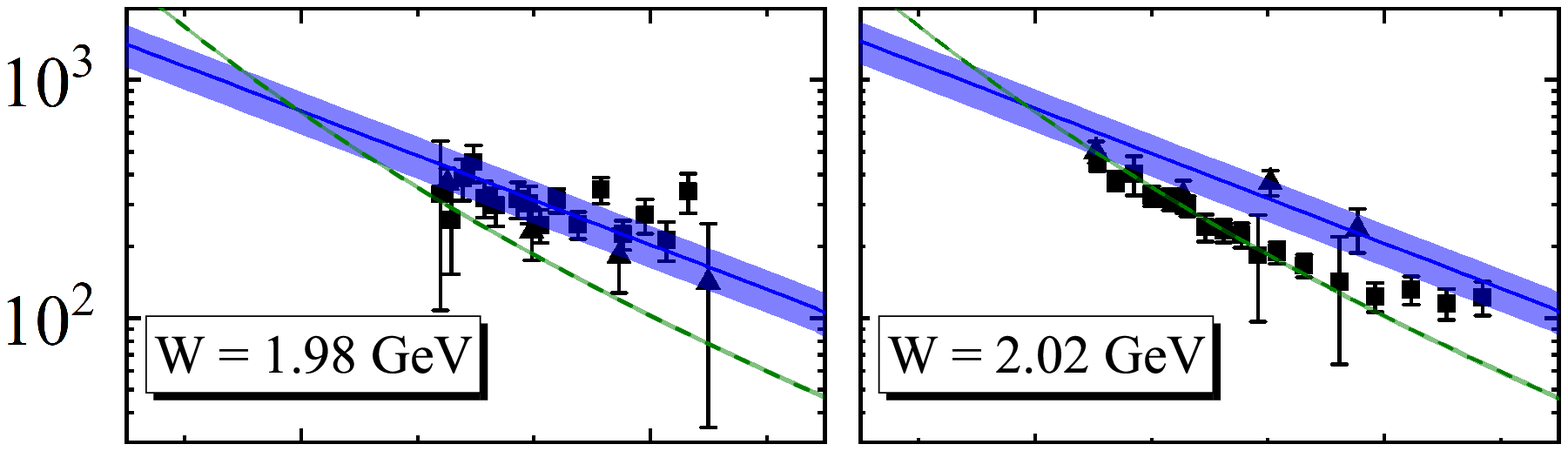}
		\\
		\includegraphics[scale=0.43]{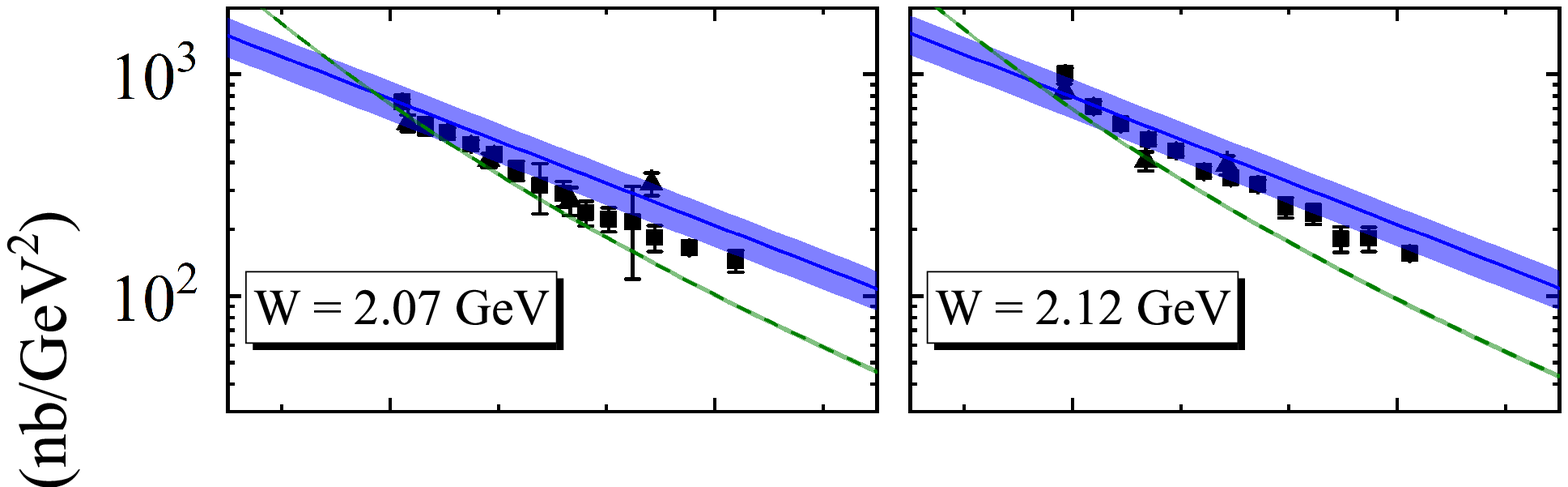}
		\\
		\includegraphics[scale=0.43]{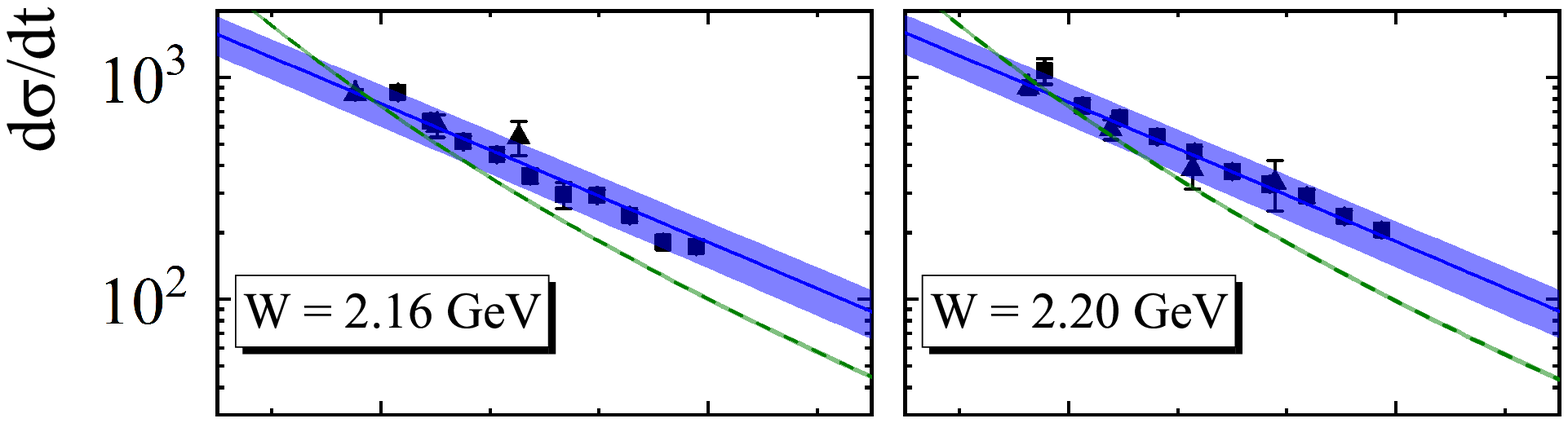}
		\\
		\includegraphics[scale=0.43]{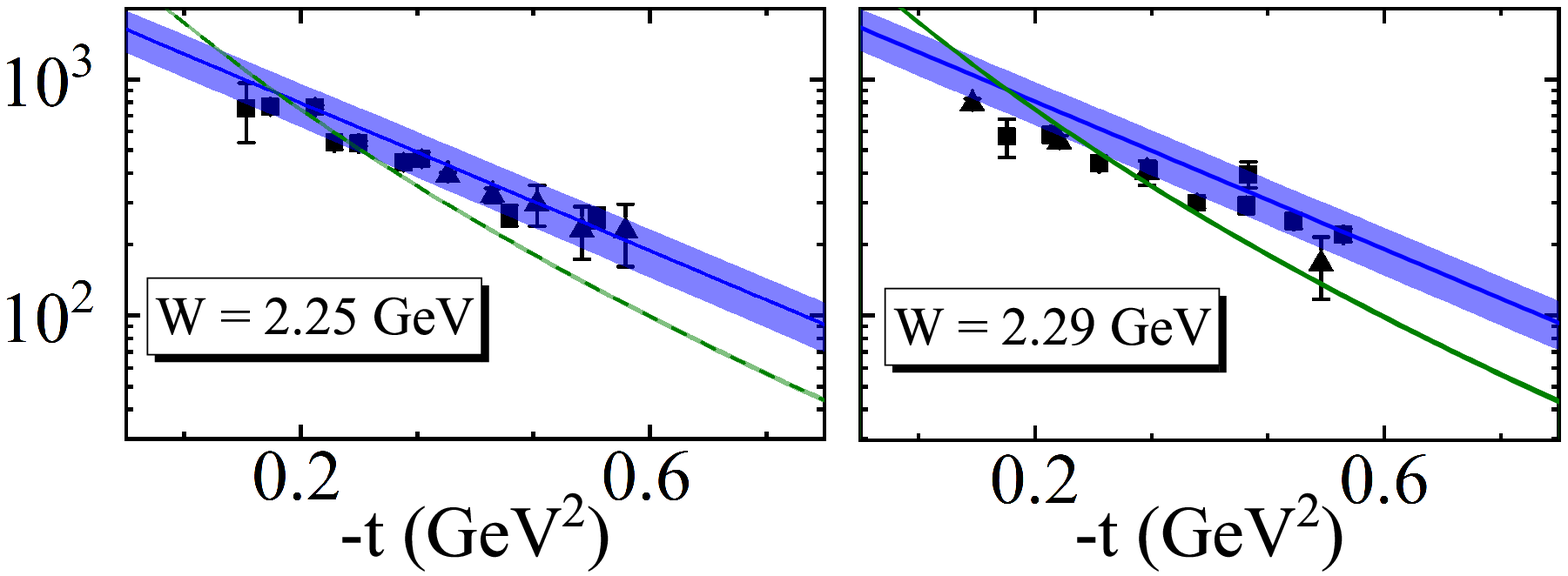}
		\\
		\includegraphics[scale=0.43]{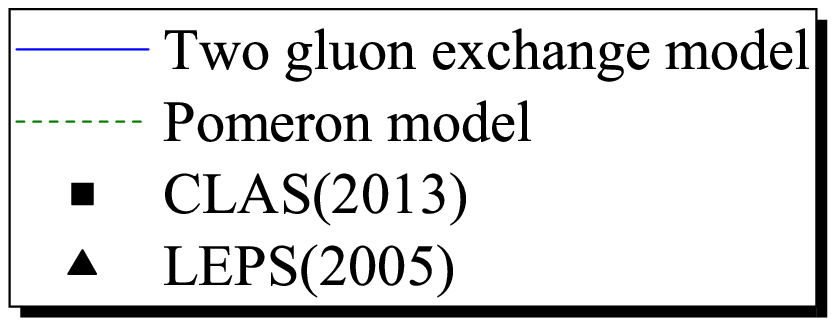}
		\caption{ The differential cross sections of the channel $\gamma p \rightarrow \phi p$ as a function of $-t$ at different $W$ values. Here, the notations are the same as in Fig. \ref{fig:tot}.}\label{fig:weifen}
	\end{center}	
\end{figure}
Fig. \ref{fig:tot} and Fig. \ref{fig:weifen} show the total and differential cross sections of $\phi$ photoproduction predicted by the two gluon exchange model and the pomeron model at the near-threshold, which are in good agreement with the experimental data of $\phi$. Based on the two models, we consider the cross section of $\phi$ photoproduction can be reliably predicted.

\begin{table} \small
	\caption{\label{tab:table3} 
	The relevant parameters for the two gluon exchange model are in the first row, and the second row is for the pomeron model.} 
	\begin{ruledtabular}
		\begin{tabular}{ccccc}
			$A_{0}$&$A_{1}$&$A_{2}$&$b_{0}$ (GeV$^{-2}$) &$\chi^{2}/$d.o.f\\
			\hline
			$0.36 \pm 0.04$&$-0.055 \pm 0.003$ & $0.12 \pm 0.03$ &$3.60\pm0.04$ &$2.87$\\
			\hline
			\hline
			$\beta$ (GeV$^{-1}$)&-&-&-&$\chi^{2}/$d.o.f\\
			\hline
			$1.919\pm0.011$&-&-&-&$9.88$
		\end{tabular}
	\end{ruledtabular}
\end{table}

\begin{figure}[htbp]
	\begin{center}
		\includegraphics[scale=0.4]{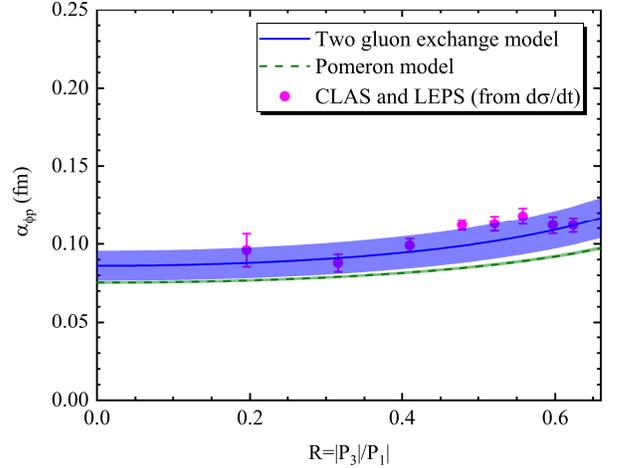}
		\caption{ The function of $|\alpha_{\phi p}|$ with $R$ from the differential experimental cross section. The blue-line is the result derived from the two gluon exchange model and the olive green dashed-line shows the result based on the pomeron model. The magenta circle is extracted directly from CLAS \cite{Dey:2014tfa} and LEPS \cite{LEPS:2005hax} data. }  \label{fig:1}
	\end{center}
\end{figure}

The scattering length $|\alpha_{\phi p}|$ obtained from the differential cross section of $\phi$ photoproduction has a slow upward trend with $R$, which can be observed in Fig. \ref{fig:1}. Here, $R$ interval is selected as $[0,0.66]$. The blue line is the $|\alpha_{\phi p}|$ based on the two gluon exchange model, with an average scattering length of $\sqrt{\left\langle |\alpha^{2}_{\phi p}|\right\rangle}=0.102\pm0.011$ fm. The olive-green dashed line is the $|\alpha_{\phi p}|$ obtained from the pomeron model with $\sqrt{\left\langle |\alpha^{2}_{\phi p}|\right\rangle}=0.087\pm0.001$ fm, which is smaller than the result from the two gluon exchange model. The magenta circle represents the $|\alpha_{\phi p}|$ extracted directly from the CLAS \cite{Dey:2014tfa} and LEPS \cite{LEPS:2005hax} data. The $|\alpha_{\phi p}|$ corresponding to each center of mass energy $W$ is listed in Tab. \ref{tab:table1}, and the $\sqrt{\left\langle |\alpha^{2}_{\phi p}|\right\rangle}=0.106\pm0.005$ fm. On the whole, the $|\alpha_{\phi p}|$ obtained from the pomeron model is not in good agreement with that extracted directly from the experiment. In contrast, the $|\alpha_{\phi p}|$ based on the two gluon exchange model agrees well.
The situation indicates that the scattering length derived from the differential cross section predicted by the two gluon exchange model is reliable.

\begin{figure}[htbp]
	\begin{center}
		\includegraphics[scale=0.4]{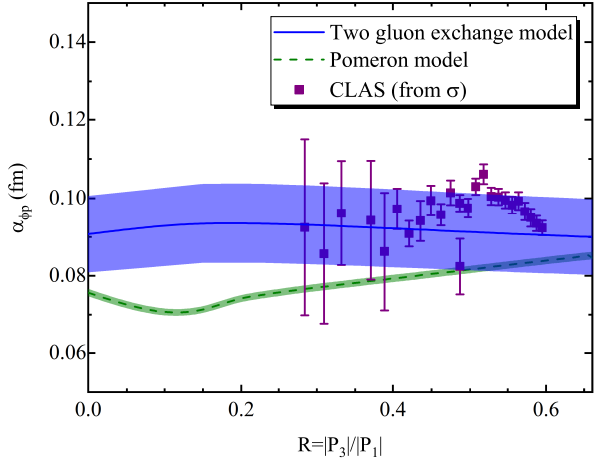}
		\caption{ The function of $|\alpha_{\phi p}|$ with $R$ via the total cross section. The purple square is the result from the CLAS \cite{Strakovsky:2020uqs}. Here, the notations are the same as in Fig. \ref{fig:1}.}  \label{fig:2}
	\end{center}
\end{figure}

As a comparison, the $|\alpha_{\phi p}|$ also can be calculated by the total cross section of $\phi$ photoproduction. The $|\alpha_{\phi p}|$ as a function of $R\in[0,0.66]$ directly calculated from the CLAS data \cite{Strakovsky:2020uqs} and based on the total cross sections predicted by the two models is shown in Fig. \ref{fig:2}. The $\sqrt{\left\langle |\alpha^{2}_{\phi p}|\right\rangle}=0.091\pm0.010$ fm based on the two gluon exchange model and $\sqrt{\left\langle |\alpha^{2}_{\phi p}|\right\rangle}=0.081\pm0.001$ fm for the pomeron model. The $\sqrt{\left\langle |\alpha^{2}_{\phi p}|\right\rangle}=0.096\pm0.010$ fm obtained directly from the total experiment cross section of the CLAS data \cite{Strakovsky:2020uqs}. Here, the same problem arises. The $|\alpha_{\phi p}|$ based on the pomeron model is obviously tiny and in poor agreement with that obtained directly from the CLAS data \cite{Strakovsky:2020uqs}. This indicates that the pomeron model's overall prediction consequence is insufficient. Therefore, the $|\alpha_{\phi p}|$ obtained by the pomeron model is ignored in the following discussion.

\begin{table}\small
\caption{\label{tab:table1} Scattering length is derived by the differential experiment cross section from CLAS \cite{Dey:2014tfa} and LEPS \cite{LEPS:2005hax} collaboration, the average is $0.106\pm0.005$ fm.}
	\begin{ruledtabular}
		\begin{tabular}{cccc}
			$W$ (GeV)  &$1.98$ &$2.02$&$2.07$ \\
			\hline
			$|\alpha_{\phi p}|$ (fm)&$0.096 \pm 0.011$&$0.088 \pm 0.006$&$0.099 \pm 0.004$\\
			\hline
			\hline
			 $W$ (GeV)&$2.12$&$2.16$&$2.20$\\
			 \hline
			$|\alpha_{\phi p}|$ (fm)&$0.109 \pm 0.003$&$0.113 \pm 0.005$&$0.118 \pm 0.005$\\
			\hline
			\hline
			$W$ (GeV)&$2.25$&$2.29$&-\\
			\hline
			$|\alpha_{\phi p}|$ (fm)&$0.112\pm0.005$&$0.108\pm0.005$&-
		\end{tabular}
	\end{ruledtabular}
\end{table}

\begin{table}[]
\begin{ruledtabular}
\caption{\label{tab:table2} Scattering length from the differential cross section $d \sigma/d t$ and total cross section $\sigma$ via the two gluon exchange model and experimental data \cite{Dey:2014tfa,LEPS:2005hax,Strakovsky:2020uqs}. The mean-square-root scattering length is $0.10\pm0.01$ fm.}
\begin{tabular}{c|cc}
	 \multirow{2}{*}{ Model } & \multicolumn{2}{c}{ Scattering length $(\mathrm{fm})$} \\
	\cline { 2 - 3 } & $d \sigma / d t$ & $\sigma$ \\
	\hline Two gluon exchange model & $0.102 \pm 0.011$ & $0.091 \pm 0.010$ \\
	\hline Extraction from experimental data & $0.106 \pm 0.005$ & $0.096 \pm 0.010$ \\
\end{tabular}
\end{ruledtabular}
\end{table}
\begin{figure}[htbp]
	\begin{center}
		\includegraphics[scale=0.4]{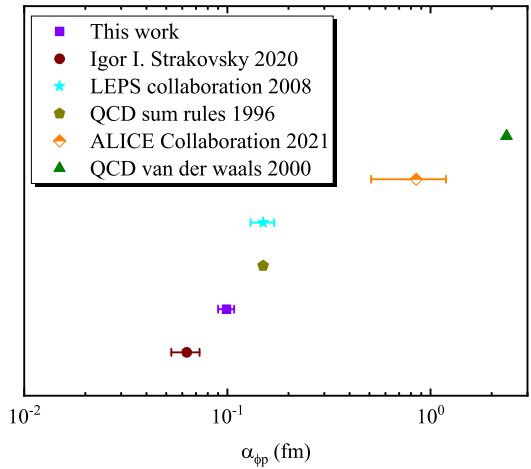}
		\caption{ The scattering length obtained by this work is compared with the absolute value obtained by other groups \cite{Strakovsky:2020uqs,Chang:2007fc,Koike:1996ga,Gao:2000az}. The purple square is $|\alpha_{\phi p}|=0.10\pm0.01$ fm. The burgundy circle is $|\alpha_{\phi p}|=0.063\pm0.01$ fm obtained by CLAS data analysis with the VMD model \cite{Strakovsky:2020uqs}.  LEPS laboratory \cite{Chang:2007fc} obtained that $|\alpha_{\phi p}|=0.15$ fm is represented by a cyan pentagram. The dark yellow pentagon is $|-0.15| \pm 0.02$ fm with the QCD sum rule \cite{Koike:1996ga}. The olive-green triangle is $|\alpha_{\phi p} |\simeq2.37$ fm by QCD van der Waals \cite{Gao:2000az}. The orange diamond is the result of the ALICE Collaboration \cite{ALICE:2021cpv}.} \label{fig:3}
	\end{center}
\end{figure}
\begin{figure}[htbp]
	\begin{center}
		\includegraphics[scale=0.4]{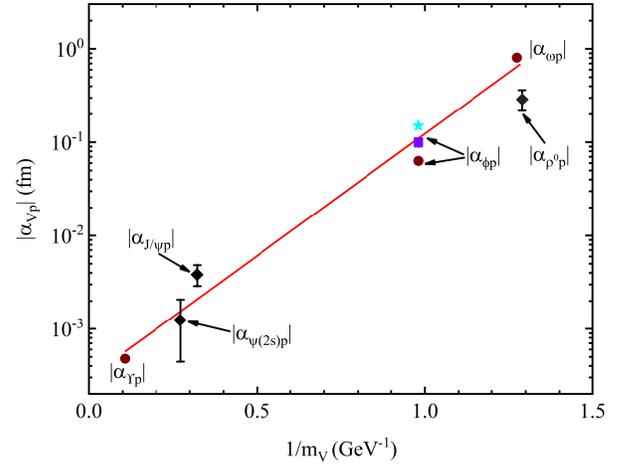}
		\caption{The scattering length of $|\alpha_{V p}|$, where $V$ is vector meson. The purple square is $|\alpha_{\phi p}|=0.10\pm0.01$ fm. The result of the black diamonds come from our previous work \cite{Wang:2022zwz,Wang:2022xpw}. The burgundy circles are the result of Ref. \cite{Strakovsky:2014wja,Strakovsky:2020uqs,Strakovsky:2021vyk}. The  cyan pentagram is from LEPS laboratory \cite{Chang:2007fc}.} \label{fig:5}
	\end{center}
\end{figure}

The above discussion shows differences in the scattering length obtained from the total and differential cross section of $\phi$ photoproduction at $R\in[0,0.66]$ which can be seen in the Tab. \ref{tab:table2}. The $\sqrt{\left\langle |\alpha^{2}_{\phi p}|\right\rangle}$ obtained directly from the experiment data \cite{Dey:2014tfa,LEPS:2005hax,Strakovsky:2020uqs} is slightly larger than that based on the two gluon exchange model, but within the error range. The $\sqrt{\left\langle |\alpha^{2}_{\phi p}|\right\rangle}$ based on the differential cross section is barely larger than that obtained from the total cross section. The reason is due to the lack of experimental data. Especially in the total cross section, the behaviour closest to the threshold scattering length cannot be observed. However, when the error bars are taken into account, the difference is completely eliminated, and the average scattering length based on the differential and total cross sections are almost identical. Therefore, we calculate the root-mean-square of the scattering length obtained from the total and differential cross sections. The final scattering length of $\phi$-$p$ interaction is $0.10\pm0.01$ fm.

A comparison of the scattering lengths of $\phi$-$p$ obtained in this work with other groups \cite{Koike:1996ga,Chang:2007fc,Koike:1996ga,Strakovsky:2020uqs,Gao:2000az,ALICE:2021cpv} is shown in Fig. \ref{fig:3}. 
Our result is roughly in the middle between that obtained by LEPS laboratory \cite{Chang:2007fc}, QCD sum rule \cite{Koike:1996ga} and Igor I. Strakovsky et al. \cite{Strakovsky:2020uqs}.
Particular attention should be paid to the result of the QCD sum rule \cite{Koike:1996ga}, which obtained $\alpha_{\phi p}=-0.15\pm0.02$ fm by introducing the vector-current nucleon forward scattering amplitude relationship without any experimental data. 
Igor I. Strakovsky et al. \cite{Strakovsky:2020uqs} also used VMD model to connect $\sigma$ with scattering length and obtained $|\alpha_{\phi p}|=0.063\pm0.012$ fm by fitting odd power. However, two relatively large results clearly go beyond our normal understanding of the scattering length of $\phi$-$p$. A result from QCD van der Waals \cite{Gao:2000az} is $|\alpha_{\phi p} |\simeq2.37$ fm and a real part of the scattering length of $\phi$-$p$ interaction calculated by ALICE Collaboration \cite{ALICE:2021cpv} from the cross section of the high-multiplicity $pp$ collisions is $0.85\pm0.34$ fm.
We suppose there should exist different reactions, so these two results are ignored for the time being.

In our previous works \cite{Wang:2022zwz,Wang:2022xpw}, the scattering lengths of $J/\psi$-$p$, $\psi(2S)$-$p$ and $\rho$-$p$ have been systematically studied. In addition, the relative vector mesons scattering length were calculated by others \cite{Strakovsky:2014wja,Chang:2007fc,Koike:1996ga,Strakovsky:2020uqs,Strakovsky:2021vyk}. The scattering lengths calculated of vector mesons with the proton interaction by different groups as shown in Fig. \ref{fig:5}. The red line represents the proportional relationship between the scattering length $|\alpha_{Vp}|$ and $\exp{(1/m_{V})}$, and the $|\alpha_{\phi p}|=0.10\pm0.01$ fm of this work intersects the red line, demonstrating that our analysis is reliable.
\subsection{The scattering length of $\phi$-$d$ interaction}
\begin{figure}[htbp]
	\begin{center}
		\includegraphics[scale=0.43]{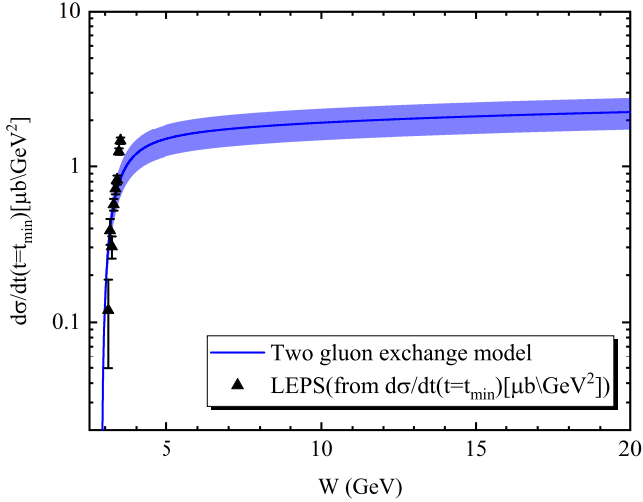}
		\caption{The $d \sigma/dt(t=t_{min})$  of $\phi$ photoproduction from deuterons as a function
        of center of mass energy $W$. The solid-line (blue) for the two gluon exchange model and the black triangle is from the LEPS Collaboration \cite{Chang:2007fc}. }  \label{fig:mxtot}
	\end{center}
\end{figure}

\begin{figure}[htbp]
	\begin{center}
		\includegraphics[scale=0.43]{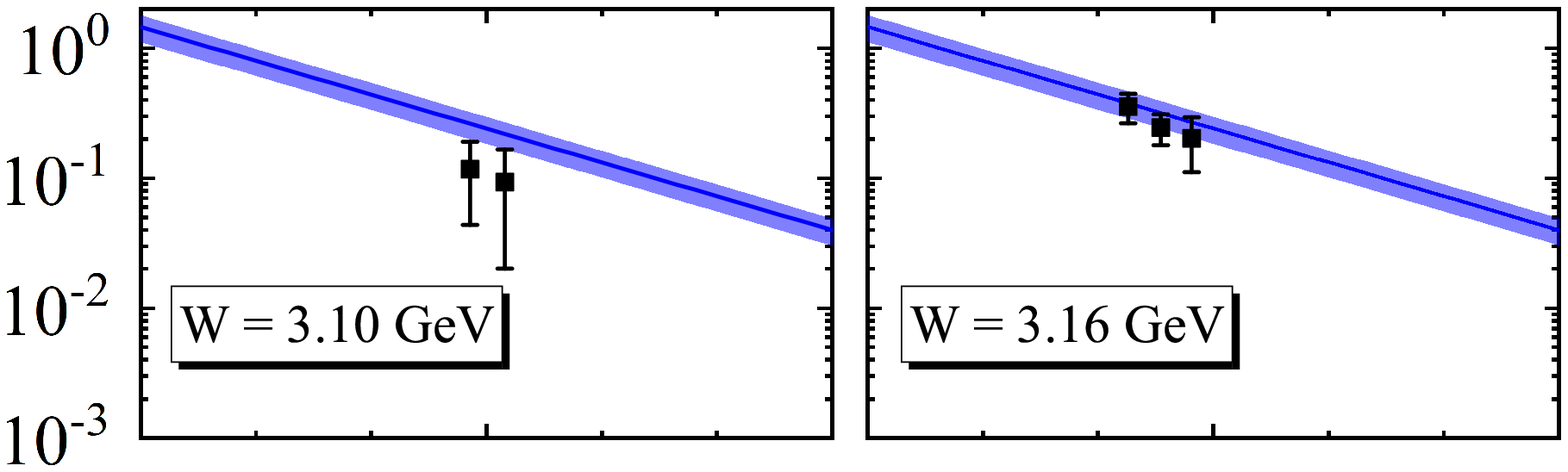}
		\\
		\includegraphics[scale=0.43]{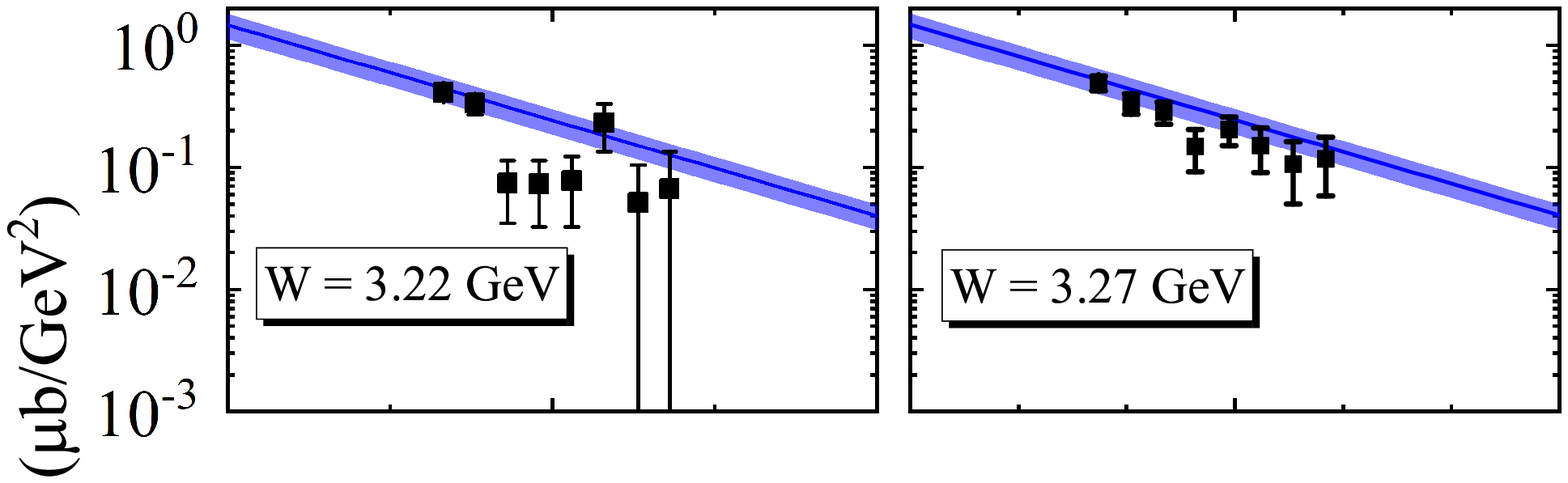}
		\\
		\includegraphics[scale=0.43]{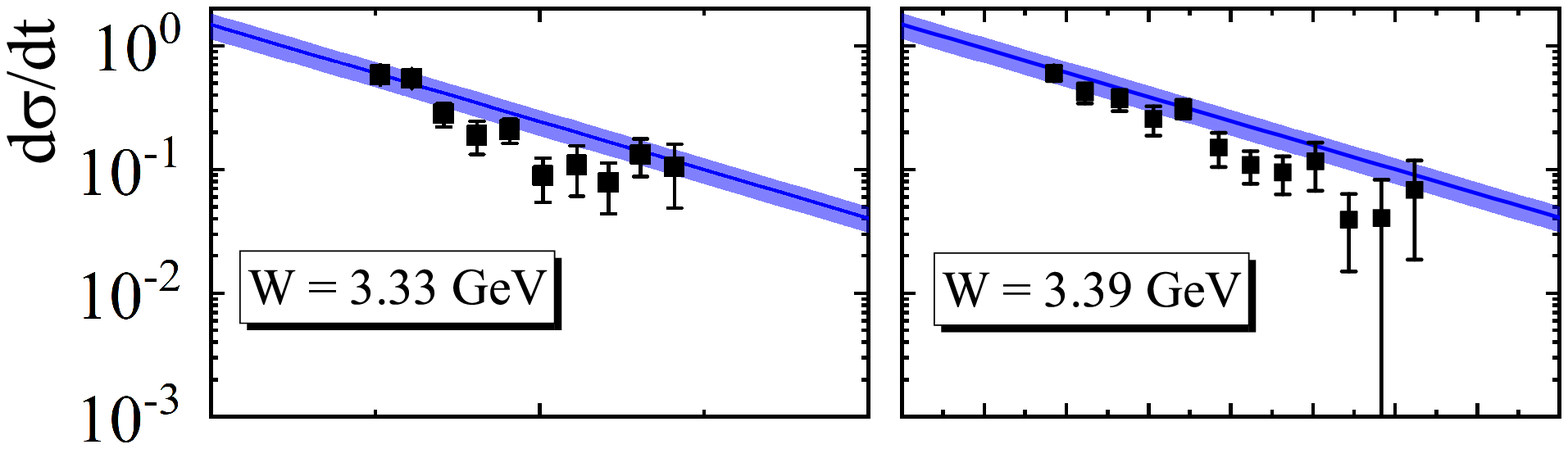}
		\\
		\includegraphics[scale=0.43]{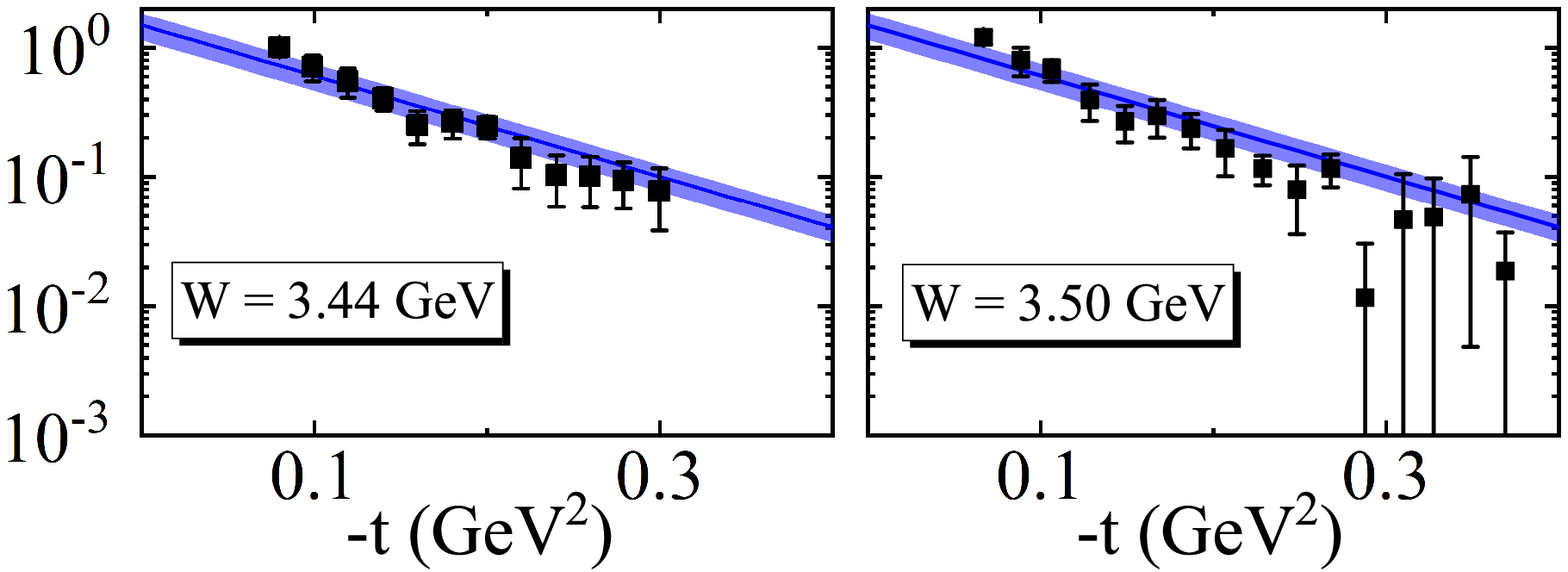}
		\\
		\includegraphics[scale=0.43]{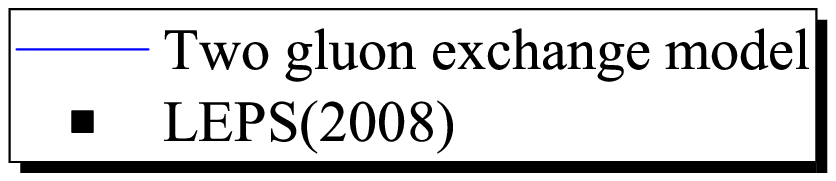}
		\caption{ The differential cross sections of $\phi$ based on the two gluon exchange model (blue line) as a function of $-t$ at different $W$ values. Here, the black square is the experimental data of $\gamma d \rightarrow \phi d$ from LEPS Collaboration \cite{Chang:2007fc}.}\label{fig:mxweifen}
	\end{center}	
\end{figure}

\begin{figure}[h]
	\begin{center}
		\includegraphics[scale=0.43]{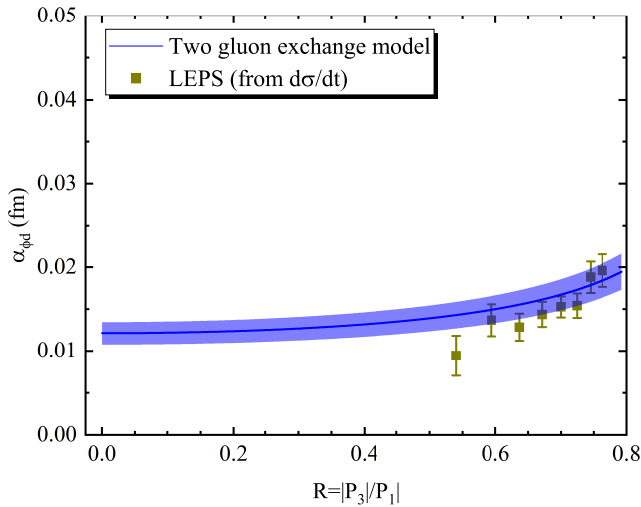}
		\caption{The function of $|\alpha_{\phi d}|$ with $R$ based on the two gluon exchange model (the blue line). The dark-yellow square is the result from LEPS \cite{Chang:2007fc} data. }  \label{fig:mxsl}
	\end{center}
\end{figure}

The analysis found that the slope of the cross section distribution of $\gamma d\rightarrow \phi d$ is larger than that of $\gamma p \rightarrow \phi p$, so it is inappropriate to describe $\phi$-$d$ based on the two gluon exchange model with $b_{0}$ =$3.6$ GeV$^{2}$. The gluon distribution function $xg(x,m_{\phi}^{2})$ obtained from $\phi$-$p$ is preserved. That is, $A_{0}$, $A_{1}$ and $A_{2}$ are retained, while $b_{0}$ is set as a free parameter and re-fitted by LEPS \cite{Chang:2007fc} data. The relevant fitting results are shown in Fig. \ref{fig:mxtot} and Fig. \ref{fig:mxweifen} which are in good agreement with the LEPS data, and 
the parameters are shown in Tab. \ref{tab:table5}. The differential cross sections of $\phi$ photoproduction from deuterons from the LEPS data are distributed at $W\in[3.10,3.50]$ GeV, corresponding to $R\in[0.54,0.76]$.
According to Eq. (\ref{eq:9}), the scattering length $|\alpha_{\phi d}|$ as a function of $R$ is shown in Fig. \ref{fig:mxsl}.
The $|\alpha_{\phi d}|$ obtained directly from the LEPS data are shown in Tab. \ref{tab:table4}, and the average scattering length $\sqrt{\left\langle |\alpha^{2}_{\phi d}|\right\rangle}=0.015\pm0.002$ fm.
At $R\in[0.54,0.76]$, the $\sqrt{\left\langle |\alpha^{2}_{\phi d}|\right\rangle}=0.016\pm0.002$ fm from two gluon exchange model. The two average scattering lengths are extremely adjacent, with a difference of only $0.001$ fm, which indicates that the $|\alpha_{\phi d}|$ obtained based on the model is reliable. 
Based on this, the average scattering length at $R\in[0,0.66]$ is calculated to be $0.014\pm0.002$ by the two gluon exchange model. Here, $R\in[0,0.66]$ is selected for $\phi$-$d$ in order to be consistent with $\phi$-$p$.


\begin{table} \small
	\caption{\label{tab:table5} 
	The  parameters $A_{0}$, $A_{1}$, $A_{2}$, $b_{0}$ and $\chi^{2} /$ d.o.f are for the process of $\gamma d \rightarrow \phi d$.} 
	\begin{ruledtabular}
		\begin{tabular}{ccccc}
			$A_{0}$&$A_{1}$&$A_{2}$&$b_{0}$ (GeV$^{-2}$)&$\chi^{2} /$ d.o.f	\\
			\hline
			$0.36 \pm 0.04$&$-0.055 \pm 0.003$ & $0.12 \pm 0.03$ &$9\pm0.29$& $0.17$\\
		\end{tabular}
	\end{ruledtabular}
\end{table}
              
\begin{figure}[t]
	\begin{center}
		\includegraphics[scale=0.43]{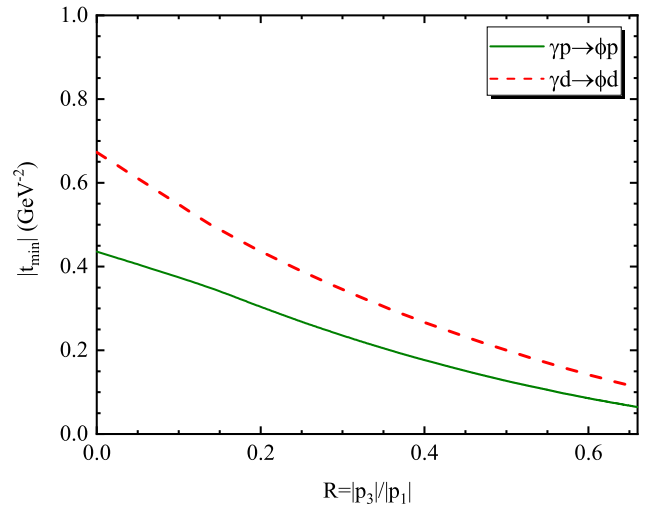}
		\caption{ The function of $|t_{min}|$ with $R$. The solid olive green line is for $\phi$-$p$ and the red dashed-line is for $\phi$-$d$. }  \label{fig:tmin}
	\end{center}
\end{figure}

According to the present result, $|\alpha_{\phi d}|$ is relatively smaller than the $|\alpha_{\phi p}|$
It can be observed from Fig. \ref{fig:weifen} and \ref{fig:mxweifen} that the differential cross sections of $\phi$-$d$ and $\phi$-$p$ photoproduction belong to the same order of magnitude, and there exists a slight distinction between them. However, the influence of $|t_{min}|$ on the differential cross section is enormous due to the diverse nucleons or nuclei interacting with $\phi$. Fig. \ref{fig:tmin} comprehensively demonstrates the difference. The $|t_{min}|$ of $\phi$-$d$ is larger than that of $\phi$-$p$ at $R\in[0,0.66]$. Also, consider that the slope of the differential cross section of $\phi$-$d$ is steeper. So $|\alpha_{\phi d}|$ $<$ $|\alpha_{\phi p}|$ is adequately explained.

\begin{table}\small
\caption{\label{tab:table4} Scattering length of $\phi$-$d$ is derived by the differential experiment cross section from LEPS \cite{Chang:2007fc} data, the average is $0.015\pm0.002$ fm.}
	\begin{ruledtabular}
		\begin{tabular}{cccc}
			$W$ (GeV)  &$3.10$ &$3.16$&$3.22$ \\
			\hline
			$|\alpha_{\phi d}|$ (fm)&$0.009 \pm 0.002$&$0.014 \pm 0.002$&$0.013 \pm 0.002$\\
			\hline
			\hline
			 $W$ (GeV)&$3.28$&$3.33$&$3.39$\\
			 \hline
			$|\alpha_{\phi d}|$ (fm)&$0.014 \pm 0.001$&$0.015 \pm 0.001$&$0.015 \pm 0.001$\\
			\hline
			\hline
			$W$ (GeV)&$3.44$&$3.50$&-\\
			\hline
			$|\alpha_{\phi d}|$ (fm)&$0.019\pm0.002$&$0.020\pm0.002$&-
		\end{tabular}
	\end{ruledtabular}
\end{table}

\section{Conclusion} \label{sec4}
In this work, based on the experimental data of $\phi$ photoproduction at the near threshold \cite{Dey:2014tfa,LEPS:2005hax,Strakovsky:2020uqs}, the interaction between the $\phi$ meson and nucleus is systematically studied under the framework of the VMD model. Specifically, the average scattering length of $\phi$-$p$ interaction is calculated to be $0.10\pm0.01$ fm which satisfies the $|\alpha_{Vp}|$ proportional to $\exp{(1/m_{V})}$. Note that our conclusions are very close to the results given by the VMD model \cite{Strakovsky:2020uqs,Chang:2007fc} and the QCD sum rule \cite{Koike:1996ga}, but are quite different from the result measured by the ALICE Collaboration \cite{ALICE:2021cpv}. This may be due to different reaction processes, in which ALICE Collaboration measures a complete two-body $\phi$-$N$ interaction, while the results obtained by VMD model may be refer to the properties of $\phi$ embedding into nucleons \cite{ALICE:2021cpv}.

Moreover, the scattering length of the $\phi$ and deuteron interaction is investigated for the first time. And the average scattering length $|\alpha_{\phi d}|$ is calculated to be $0.014\pm0.002$ fm, which is nearly seven times smaller than that of $|\alpha_{\phi p}|$. This result indicates that the coupling between $\phi$-$d$ is much weaker than that between $\phi$-$p$. One possible explanation is that since the deuteron is composed of two nucleons, there is an overall effect that the $\phi$ meson is more fully embedded in the deuteron than in the nucleon \cite{ALICE:2021cpv}, making the $\phi$-$d$ scattering length smaller. Of course, this is still an open question and more explanation and discussion is needed.

It should be mentioned that although the pomeron model and the parameterized two gluon exchange model can describe the vector meson photoproduction data well, in fact, there are usually contributions from intermediate exchange mesons or nucleon resonances in the low energy region. For example, in Refs. \cite{Titov:1999eu,Lebiedowicz:2019boz}, the contribution of $\pi^{0}$ and $\eta$ meson exchange to the cross section of vector meson photoproduction was discussed. In Refs. \cite{Kiswandhi:2011cq,Kim:2019kef,Ozaki:2009mj,Ryu:2012tw}, the role of the intermediate nucleon resonances and the direct $\phi$ meson radiation via proton exchanges has also been studied. These results indicate that more research on the production mechanism of vector meson photoproduction at low energy is still needed, which is helpful to more accurately explain the interaction between vector meson and nucleus.

In order to study these issues more accurately, it is necessary to measure more accurate experimental data of vector mesons photoproduction or electroproduction. At present, the EIC facilities \cite{Accardi:2012qut,Anderle:2021wcy} in China and the United States have listed nuclear structure and properties as an important scientific objectives, and $\phi$-$N$ interaction is one of the essential directions. Our result can provide a crucial reference and basis for detecting the scattering length of $\phi$-$N$ interaction more accurately in the future.
\begin{acknowledgments}
This work is supported
by the National Natural Science Foundation of China under
Grant Nos. 12065014 and 12047501, and by the the Natural Science
Foundation of Gansu province under Grant No. 22JR5RA266. We acknowledge the West Light Foundation of The Chinese Academy of Sciences, Grant No.
21JR7RA201.
\end{acknowledgments}


\begin{thebibliography}{99}
\bibitem{Titov:2007xb}
A.~I.~Titov, T.~Nakano, S.~Date and Y.~Ohashi,
``Comments on differential cross-section of phi-meson photoproduction at threshold,''
Phys. Rev. C \textbf{76}, 048202 (2007).

\bibitem{Strakovsky:2020uqs}
I.~I.~Strakovsky, L.~Pentchev and A.~Titov,
``Comparative analysis of $\omega p$, $\phi p$, and $J/\psi p$ scattering lengths from A2, CLAS, and GlueX threshold measurements,''
Phys. Rev. C \textbf{101}, 045201 (2020).

\bibitem{Chang:2007fc}
W.~C.~Chang, K.~Horie, S.~Shimizu, M.~Miyabe, D.~S.~Ahn, J.~K.~Ahn, H.~Akimune, Y.~Asano, S.~Date and H.~Ejiri, \textit{et al.}
``Forward coherent phi-meson photoproduction from deuterons near threshold,''
Phys. Lett. B \textbf{658}, 209-215 (2008).

\bibitem{Koike:1996ga}
Y.~Koike and A.~Hayashigaki,
``QCD sum rules for rho, omega, phi meson - nucleon scattering lengths and the mass shifts in nuclear medium,''
Prog. Theor. Phys. \textbf{98}, 631-652 (1997).

\bibitem{Wang:2022zwz}
X.~Y.~Wang, F.~Zeng, Q.~Wang and L.~Zhang,
``First extraction of proton mass radius and scattering length $\left|\alpha_{\rho^0 p}\right|$ from $\rho^0$ photoproduction,''
[arXiv:2206.09170 [nucl-th]].

\bibitem{Wang:2022xpw}
X.~Y.~Wang, F.~Zeng and I.~I.~Strakovsky,
``The $\psi^{(\ast)}p$ scattering length based on near-threshold charmoniums photoproduction,''
Phys. Rev. C \textbf{106}, 015202 (2022).

\bibitem{Strakovsky:2014wja}
I.~I.~Strakovsky, S.~Prakhov, Y.~I.~Azimov, P.~Aguar-Bartolom\'e, J.~R.~M.~Annand, H.~J.~Arends, K.~Bantawa, R.~Beck, V.~Bekrenev and H.~Bergh\"auser, \textit{et al.}
``Photoproduction of the \ensuremath{\omega} meson on the proton near threshold,''
Phys. Rev. C \textbf{91}, no.4, 045207 (2015).

\bibitem{Strakovsky:2021vyk}
I.~I.~Strakovsky, W.~J.~Briscoe, L.~Pentchev and A.~Schmidt,
``Threshold Upsilon-meson photoproduction at the EIC and EicC,''
Phys. Rev. D \textbf{104}, 074028 (2021).

\bibitem{Pentchev:2020kao}
L.~Pentchev and I.~I.~Strakovsky,
``$J/\psi$-$p$ Scattering Length from the Total and Differential Photoproduction Cross Sections,''
Eur. Phys. J. A \textbf{57}, 56 (2021).

\bibitem{Strakovsky:2019bev}
I.~Strakovsky, D.~Epifanov and L.~Pentchev,
``J/$\psi$p scattering length from GlueX threshold measurements,''
Phys. Rev. C \textbf{101}, 042201 (2020).

\bibitem{Wang:2021ujy}
R.~Wang, W.~Kou, C.~Han, J.~Evslin and X.~Chen,
``Proton and deuteron mass radii from near-threshold \ensuremath{\phi}-meson photoproduction,''
Phys. Rev. D \textbf{104}, 074033 (2021).

\bibitem{Wang:2022uch}
X.~Y.~Wang, C.~Dong and Q.~Wang,
``Research on proton mass radius and its mechanical properties via strange $\phi$ meson photoproduction,''
Phys. Rev. D 106, 056027.

\bibitem{Gao:2000az}
H.~Gao, T.~S.~H.~Lee and V.~Marinov,
``Phi0 - N bound state,''
Phys. Rev. C \textbf{63}, 022201 (2001).

\bibitem{GlueX:2019mkq}
A.~Ali \textit{et al.} [GlueX],
``First Measurement of Near-Threshold J/\ensuremath{\psi} Exclusive Photoproduction off the Proton,''
Phys. Rev. Lett. \textbf{123}, 072001 (2019).

\bibitem{Gryniuk:2016mpk}
O.~Gryniuk and M.~Vanderhaeghen,
``Accessing the real part of the forward $J/\psi$-p scattering amplitude from $J/\psi$ photoproduction on protons around threshold,''
Phys. Rev. D \textbf{94}, 074001 (2016)

\bibitem{ALICE:2021cpv}
S.~Acharya \textit{et al.} [ALICE],
``Experimental Evidence for an Attractive p-$\phi$ Interaction,''
Phys. Rev. Lett. \textbf{127}, 172301 (2021).

\bibitem{Sibirtsev:2004ca}
A.~Sibirtsev, S.~Krewald and A.~W.~Thomas,
``Systematic analysis of charmonium photoproduction,''
J. Phys. G \textbf{30}, 1427-1444 (2004).

\bibitem{Laget:1994ba}
J.~M.~Laget and R.~Mendez-Galain,
``Exclusive photoproduction and electroproduction of vector mesons at large momentum transfer,''
Nucl. Phys. A \textbf{581}, 397-428 (1995).

\bibitem{Dey:2014tfa}
B.~Dey \textit{et al.} [CLAS],
``Data analysis techniques, differential cross sections, and spin density matrix elements for the reaction $\gamma p \rightarrow \phi p$,''
Phys. Rev. C \textbf{89}, 055208 (2014).

\bibitem{LEPS:2005hax}
T.~Mibe \textit{et al.} [LEPS],
``Diffractive $\phi$-meson photoproduction on proton near threshold,''
Phys. Rev. Lett. \textbf{95}, 182001 (2005).

\bibitem{VMD}
Eisenberg, Y.  and  Haber, B.  and  Kogan, E.  and  Ronat, E. E.  and  Shapira, A.  and  Yekutieli, G.
``Photoproduction of $\rho^{0}$ and $\omega$ in $\gamma d$ interactions at 4.3 GEV,''
Nuclear Physics B42 (1972) 349-368. 

\bibitem{Gell-Mann:1961jim}
M.~Gell-Mann and F.~Zachariasen,
``Form-factors and vector mesons,''
Phys. Rev. \textbf{124}, 953-964 (1961).

\bibitem{Harari:1966ms}
H.~Harari,
``Phenomenological analysis of the photoproduction of neutral vector mesons and strange particles,''
Phys. Rev. \textbf{155}, 1565-1578 (1967).

\bibitem{Stodolsky:1966am}
L.~Stodolsky,
``Hadron-like behavior of gamma, neutrino nuclear cross-sections,''
Phys. Rev. Lett. \textbf{18}, 135-137 (1967).

\bibitem{Dar:1968sla}
A.~Dar, V.~F.~Weisskopf, C.~Levinson and H.~Lipkin,
``Vector Dominance in Photoproduction,''
Phys. Rev. Lett. \textbf{20}, 1261-1265 (1968).

\bibitem{Accardi:2012qut}
A.~Accardi, J.~L.~Albacete, M.~Anselmino, N.~Armesto, E.~C.~Aschenauer, A.~Bacchetta, D.~Boer, W.~K.~Brooks, T.~Burton and N.~B.~Chang, \textit{et al.}
``Electron Ion Collider: The Next QCD Frontier: Understanding the glue that binds us all,''
Eur. Phys. J. A \textbf{52}, 268 (2016).

\bibitem{Anderle:2021wcy}
D.~P.~Anderle, V.~Bertone, X.~Cao, L.~Chang, N.~Chang, G.~Chen, X.~Chen, Z.~Chen, Z.~Cui and L.~Dai, \textit{et al.}
``Electron-ion collider in China,''
Front. Phys. (Beijing) \textbf{16}, 64701 (2021).

\bibitem{Kou:2021bez}
W.~Kou, R.~Wang and X.~Chen,
``Determination of the Proton Trace Anomaly Energy from the Near-Threshold Vector Meson Photoproduction Data,''
[arXiv:2103.10017 [hep-ph]].

\bibitem{Landshoff:1990kj}
P.~V.~Landshoff,
``Nonperturbative effects at small x,''
Nucl. Phys. B Proc. Suppl. \textbf{18}, 211-219 (1991).


\bibitem{Barber:1981fj}
D.~P.~Barber, J.~B.~Dainton, L.~C.~Y.~Lee, R.~Marshall, J.~C.~Thompson, D.~T.~Williams, T.~J.~Brodbeck, G.~Frost, G.~N.~Patrick and G.~F.~Pearce, \textit{et al.}
``A Study of Elastic Photoproduction of Low Mass K$^{+} $K$^{-}$ Pairs From Hydrogen in the Energy Range 2.8-{GeV} to 4.8-{GeV},''
Z. Phys. C \textbf{12}, 1 (1982).

\bibitem{Ballam:1972eq}
J.~Ballam, G.~B.~Chadwick, Y.~Eisenberg, E.~Kogan, K.~C.~Moffeit, P.~Seyboth, I.~O.~Skillicorn, H.~Spitzer, G.~E.~Wolf and H.~H.~Bingham, \textit{et al.}
``Vector Meson Production by Polarized Photons at 2.8-GeV, 4.7-GeV, and 9.3-GeV,''
Phys. Rev. D \textbf{7}, 3150 (1973).

\bibitem{Egloff:1979mg}
R.~M.~Egloff, P.~J.~Davis, G.~Luste, J.~F.~Martin, J.~D.~Prentice, D.~O.~Caldwell, J.~P.~Cumalat, A.~M.~Eisner, A.~Lu and R.~J.~Morrison, \textit{et al.}
``Measurements of Elastic Rho and Phi Meson Photoproduction Cross-Sections on Protons from 30 GeV to 180 GeV,''
Phys. Rev. Lett. \textbf{43}, 657 (1979).

\bibitem{Busenitz:1989gq}
J.~Busenitz, C.~Olszewski, P.~Callahan, G.~Gladding, A.~Wattenberg, M.~E.~Binkley, J.~Butler, J.~P.~Cumalat, I.~Gaines and M.~Gormley, \textit{et al.}
``High-energy Photoproduction of $\pi^+ \pi^- \pi^0$, $K^+ K^-$, and $P \bar{P}$ States,''
Phys. Rev. D \textbf{40}, 1-21 (1989).

\bibitem{Owens:2012bv}
J.~F.~Owens, A.~Accardi and W.~Melnitchouk,
``Global parton distributions with nuclear and finite-$Q^2$ corrections,''
Phys. Rev. D \textbf{87}, 094012 (2013).

\bibitem{Titov:1999eu}
A.~I.~Titov, T.~S.~H.~Lee, H.~Toki and O.~Streltsova,
``Structure of the fgr photoproduction amplitude at a few GeV,''
Phys. Rev. C \textbf{60}, 035205 (1999).

\bibitem{Lebiedowicz:2019boz}
P.~Lebiedowicz, O.~Nachtmann and A.~Szczurek,
``Searching for the odderon in  $pp \to pp K^{+}K^{-}$ and $pp \to pp \mu^{+}\mu^{-}$ reactions in the $\phi(1020)$ resonance region at the LHC,''
Phys. Rev. D \textbf{101}, 094012 (2020).

\bibitem{Kiswandhi:2011cq}
A.~Kiswandhi and S.~N.~Yang,
``On the near-threshold peak structure in the differential cross section of \textbackslash{}phi-meson photoproduction: indication of a missing resonance with non-negligible strangeness content,''
Phys. Rev. C \textbf{86}, 015203 (2012)
[erratum: Phys. Rev. C \textbf{86}, 019904 (2012)].

\bibitem{Kim:2019kef}
S.~H.~Kim and S.~i.~Nam,
``Pomeron, nucleon-resonance, and $(0^+,0^-,1^+)$-meson contributions in $\phi$-meson photoproduction,''
Phys. Rev. C \textbf{100}, 065208 (2019).

\bibitem{Ozaki:2009mj}
S.~Ozaki, A.~Hosaka, H.~Nagahiro and O.~Scholten,
``A Coupled-channel analysis for phi-photoproduction with Lambda(1520),''
Phys. Rev. C \textbf{80}, 035201 (2009)
[erratum: Phys. Rev. C \textbf{81}, 059901 (2010)].

\bibitem{Ryu:2012tw}
H.~Y.~Ryu, A.~I.~Titov, A.~Hosaka and H.~C.~Kim,
``$\phi$ photoprodution with coupled-channel effects,''
PTEP \textbf{2014}, 023D03 (2014).


\end{thebibliography}
\end{document}